\documentclass[a4paper,12pt]{article}
\usepackage{amssymb,latexsym}
\usepackage{amsmath}

\newcommand{\ph}{\phantom}
\newcommand{\ba}{\begin{eqnarray}}

\newcommand{\ea}{\end{eqnarray}}
\newcommand{\be}{\begin{equation}}

\newcommand{\ee}{\end{equation}}

\newcommand{\de}{\delta}

\newcommand{\mb}{\mathbf}
\newcommand{\mc}{\mathcal}
\newcommand{\bb}{\mathbb}
\newcommand{\p}{\partial}

\newcommand{\rar}{\rightarrow}

\newcommand{\nn}{\nonumber}
\newcommand{\dar}{\Downarrow}

\newcommand{\im}{\mathrm{Im}}
\newcommand{\re}{\mathrm{Re}}

\makeatletter \@addtoreset{equation}{section} \makeatother

\usepackage{fullpage}

\makeatletter \@addtoreset{equation}{section}
\makeatother

\begin{document}
\begin{titlepage}
   \thispagestyle{empty}
   \begin{flushright}
       \hfill{CERN-PH-TH/2009-150}\\
       \hfill{DFPD-09/TH/16}\\
   \end{flushright}

   \vspace{35pt}

   \begin{center}
       { \LARGE{\bf      {5d/4d U-dualities and $\mc N\!=\!8$~black~holes}}}

       \vspace{50pt}

       {\large Anna Ceresole$^a$, Sergio Ferrara $^{b,c}$ and Alessandra Gnecchi$^{d}$}

       \vspace{40pt}

       {\it ${}^a$ INFN, Sezione di Torino $\&$ Dipartimento di Fisica Teorica\\
       Universit\`a di Torino, Via Pietro Giuria 1, 10125 Torino, Italy}

       \vspace{15pt}

       {\it ${}^b$ Theory Division - CERN,\\
       CH 1211, Geneva 23, Switzerland}

       \vspace{15pt}

       {\it ${}^c$ INFN - LNF, \\
       Via Enrico Fermi 40, I-00044 Frascati, Italy}

       \vspace{15pt}

       {\it ${}^d$ Dipartimento di Fisica ``Galileo Galilei'' $\&$ INFN, Sezione di Padova \\
       Universit\`a di Padova, Via Marzolo 8, 35131 Padova, Italy}

       \vspace{15pt}

       \vspace{40pt}

       {ABSTRACT}
   \end{center}

   \vspace{10pt}
\noindent
  We use the connection between the U-duality groups in $d=5$ and $d=4$ to derive properties of the $\mc N=8$ black hole potential and its critical points (attractors).
This approach allows to study and compare the supersymmetry features of different solutions.

\end{titlepage}

\baselineskip 6 mm


\newpage\tableofcontents

\section{\label{Intro}Introduction}

The $\mc N=8$ supergravity theory in $d=4$ \cite{Cremmer:1979up} and $d=5$ \cite{Cremmer5d} dimensions is a remarkable theory which unifies the
gravitational fields with other lower spin particles in a rather unique way, due to the high constraints of local $\mc N=8$ supersymmetry, the
maximal one realized in a 4d Lagrangian field theory. These theories, particularly in four dimensions, are supposed to enjoy exceptional
ultraviolet properties. For this reason,  4d supergravity has been advocated  not only as the simplest quantum field theory \cite{Arkani} but also
as a potential candidate for a finite theory of quantum gravity, even without its completion into a larger theory \cite{Bern}. Maximal
supergravity in highest dimensions  has a large number of  classical solutions \cite{Gibbons:1993sv} which may survive at the quantum level. Among
them, there are black p-branes of several types\cite{Duff} and interestingly, 4d black holes of different nature.

On the other hand, theories with lower supersymmetries  (such as $\mc N=2$) emerging from Calabi-Yau compactifications of M-theory or superstring
theory, admit extremal black hole solutions that have been the subject of intense study, because of their wide range of classical and quantum
aspects. For asymptotically flat, stationary and spherically symmetric extremal black holes, the attractor behaviour \cite{AM,universality} has
played an important role not only in determining universal features of fields flows toward the horizon, but also to explore dynamical properties
such as wall crossing\cite{denef1} and split attractor flows\cite{denef2}, the connections with string topological partition functions\cite{OSV}
and relations with microstates counting\cite{sen} . Therefore, it has become natural to study the properties of extremal black holes  not only in
the context of $\mc N=2$, but also in theories with higher supersymmetries, up to $\mc N=8$\cite{Cvetic:1995yp}-\cite{Sen:2009gy}.

In $\mc N=8$ supergravity, in the Einsteinian approximation, there is a nice relation between the classification of large black holes which undergo the attractor flow and charge orbits which classify, in a duality invariant manner, the properties of the dyonic vector of electric and magnetic charges $Q=(p^{\Lambda},\,q_{\Lambda})$  ($\Lambda=0,...,27$ in $d=4$) \cite{malda,gunaydin1}.
The attractor points are given by extrema of the $4d$ black hole potential, which is given by \cite{revisited,Ferrara:2006em}
\begin{eqnarray}\label{}
V_{BH}&=&\frac12 Z_{AB}Z^{*\,AB}=\langle Q,V_{AB} \rangle\ \langle Q, \overline V^{AB} \rangle\ ,
\end{eqnarray}
where  the central charge is the antisymmetric matrix ($A,B=1,...,8$)
\begin{eqnarray}\label{ZAB}
Z_{AB}&=&\langle Q,V_{AB} \rangle=Q^{T}\,\Omega\, V_{AB}=f^{\Lambda}_{\ph\Lambda AB}\,q_{\Lambda}-h_{\Lambda\, AB}\,p^{\Lambda}\ ,
\end{eqnarray}
the symplectic sections are
\begin{eqnarray}\label{}
V_{AB}&=&(f^{\Lambda}_{\ph\Lambda AB},h_{\Lambda\, AB})\ ,
\end{eqnarray}
and $\Omega$ is the symplectic invariant metric.

An important role is played by the Cartan quartic invariant $I_{4}$\cite{Cartan,Cremmer:1979up} in that it only depends on $Q$ and not on the
asymptotic values of the $70$ scalar fields $\varphi$. This means that if we construct $I_{4}$ as a combination of quartic powers of the central
charge matrix $Z_{AB}(q,p,\varphi)$ \cite{Kallosh:1996uy}, the $\varphi$ dependence drops out from the final expression
\begin{eqnarray}\label{}
\frac{\p}{\p\varphi}I_{4}(Z_{AB})&=&0\ .
\end{eqnarray}
Analogue (cubic) invariants $I_{3}$ exist for black holes and/or (black) strings in $d=5$\cite{universality,malda}. These are given by
\begin{eqnarray}\label{}
I_{3}(p^{I})&=&\frac1{3!}d_{IJK}p^{I}p^{J}p^{K}\ ,\\
I_{3}(q_{I})&=&\frac1{3!}d^{IJK}q_{I}q_{J}q_{K}\ ,
\end{eqnarray}
where $d_{IJK}$, $d^{IJK}$ are the $(27)^{3}$ $E_{6(6)}$ invariants.  Consequently, the $d=4$  $E_{7(7)}$ quartic invariant takes the form
\begin{eqnarray}\label{}
I_{4}(Q)&=&-(p^{0}q_{0}+p^{I}q_{I})^{2}+4\left[ -p^{0}I_{3}(q)+q_{0}I_{3}(p) +
\frac{\p I_{3}(q)}{\p{q_{I}}}\frac{\p I_{3}(p)}{\p p^{I}} \right]\ .
\end{eqnarray}

On the other hand, in terms of  the central charge matrices $Z_{ab}(\phi,q)$ (in $d=5$ this is the $\mb{27}$ representation of $USp(8)$) and
$Z_{AB}(\phi,p,q)$ (in $d=4$ this is the $\mb{28}$ of $SU(8)$), their expression is
\begin{eqnarray}\label{}
I_{3}(q)&=& Z_{ab}\Omega^{bc}Z_{cd}\Omega^{dq}Z_{qp}\Omega^{pa}\ , \qquad Z_{ab}\Omega^{ab}=0\ ,
\\
I_{4}(p,q)&=&\frac14\left[ 4\,Tr (ZZ^{\dag}ZZ^{\dag})-(Tr\,ZZ^{\dag})^{2}+32\,\re\,(Pf\,Z_{AB})
\right]\ ,
\end{eqnarray}
where  $ZZ^{\dag}=Z_{AB}\bar Z^{CB}$, $\Omega^{ab}$ is the $5d$ symplectic invariant metric, and the Pfaffian of the central charge is
\cite{Cremmer:1979up}
\begin{eqnarray}\label{}
Pf\,(Z_{AB})=\frac{1}{2^{4}4!}\epsilon^{ABCDEFGH}Z_{AB}Z_{CD}Z_{EF}Z_{GH}\ .
\end{eqnarray}
In fact, these are simply the (totally symmetric) invariants which characterize the 27 dimensional representation of $E_{6(6)}$ and the ${56}$ dimensional representation of $E_{7(7)}$, which are the $U$-duality \cite{Hull:1994ys} symmetries of $\mc N=8$ supergravity in  $d=5$ and $d=4$, respectively.

When charges are chosen such that $I_{4}$ and $I_{3}$ are not vanishing, one has large black holes and in the extremal case the attractor behaviour may occur.  However, while at $d=5$ there is a unique ($\frac18$-BPS) attractor orbit with $I_{3}\neq0$, associated to the space\cite{gunaydin1,Lu:1997bg}
\begin{eqnarray}\label{}
\mc O_{d=5}&=&\frac{E_{6(6)}}{F_{4(4)}}\ ,
\end{eqnarray}
at $d=4$ two orbits emerge, the BPS one
\begin{eqnarray}\label{}
\mc O_{d=4,\,BPS} &=&\frac{E_{7(7)}}{E_{6(2)}}\ ,
\end{eqnarray}
and the non BPS one with different stabilizer
\begin{eqnarray}\label{}
\mc O_{d=4,\,non-BPS}&=&\frac{E_{7(7)}}{E_{6(6)}}\ .
\end{eqnarray}
Such orbits have further ramifications in theories with lower supersymmetry , but it is the aim of this paper to confine our attention to the $\mc N=8$ theory.

In this paper, extending a previous result for $\mc N=2$ theories \cite{Ceresole:2007rq}, we elucidate the connection between these configurations and  we relate the critical points of the $\mc N=8$ black hole potential of the $5d$ and $4d$ theories.
To achieve this goal we use a formulation of $4d$ supergravity in a $E_{6(6)}$ duality covariant basis \cite{ADFL}, which is appropriate to discuss a $4d/5d$ correspondence. This is not the same as the Cremmer-Julia\cite{Cremmer:1979up} or de Wit-Nicolai\cite{dewitnicolai} manifest $SO(8)$ (and $SL(8,\bb R)$) covariant formulation, but it is rather related to the Sezgin-Van Nieuwenhuizen $5d/4d$ dimensional reduction\cite{Sezgin:1981ac}. These two formulations are related to one another by dualizing several of the vector fields and therefore they interchange electric and magnetic charges of some of the 28 vector fields of the final theory. The precise relation between these theories was recently discussed in \cite{Ceresole:2009jc}.

The paper is organized as follows.
In sec. \ref{VBHrewritten} we rewrite the $4d$ black hole potential in terms of  central charges. This is essential in order to discuss the supersymmetry properties of the solutions. In fact, in the specific solutions we consider in sec. \ref{solutions-5d} and \ref{solutions-4d}, BPS and non-BPS critical points are simply obtained by some charges sign flip.  This will manifest in completely different symmetry properties of the central charge matrix, in the normal frame, at the fixed point. These properties reflect the different character of the BPS and non BPS  charge orbits.

The solutions of the critical point equations are particularly simple in the  ``axion free'' case, discussed in sec. \ref{solutions-5d} and \ref{solutions-4d}, which only occur for some chosen charge configurations.   In sec. \ref{solutions-5d} we derive critical point equations that are completely general and that  may be used to study any solution.

The formula for the  $\mc N=8$ potential  given in sec.\ \ref{VBHrewritten} was obtained in an earlier work \cite{Ceresole:2009jc}, and it is identical to the $\mc N=2$ case \cite{Ceresole:2007rq}. The only difference relies in the  kinetic matrix $a_{IJ}$ which, in $\mc N=2$ is given by real special geometry while in $\mc N=8$ is given in terms of the $E_{6(6)}$ coset representatives \cite{Sezgin:1981ac,revisited}. However, in the normal frame, when we suitably restrict to two moduli, this matrix does indeed become an $\mc N=2$ matrix, although the interpretation in terms of central charges is completely different.

The supersymmetry properties of the solutions in the $\mc N=8$ and $\mc N=2$ theories are compared in subsection \ref{N8N2}.
We will see that in the $\mc N=2$ interpretation, depending on the sign  of the charges, both a BPS and a non-BPS branch exist in $d=5$ while  two non BPS branches exist in the $d=4$ theory. In  $\mc N=8$, the occurrence of  one less branch in both dimensions is due to the fact that the central and matter charges of the $\mc N=2$ theory are all  embedded in the central charge matrix of the $\mc N=8$ theory. The higher number of attractive orbits can also be explained by the different form of the relevant non compact groups and their stabilizers for the moduli space of solutions.

\section{\label{VBHrewritten}4d/5d relations for the $\mc N=8$ extremal black hole potential}

In this section we remind the reader how the $\mc N=8$ potential was derived in a basis that illustrates the relation between $4$ and $5$ dimensions \cite{Ceresole:2009jc}.

Using known identities \cite{Ferrara:2006em,CDF}, the black hole potential can be written as a quadratic form in terms of the charge vector $Q$
and the symplectic $56\times56$ matrix $\mc M(\mc N)$, related to the $4d$ vector kinetic matrix $\mc N_{\Lambda\Sigma}$
\ba\label {Vbh} V_{BH}&=&-\frac12Q^{T}\mc M(\mc
N)Q\, , \ea
where $\mc M$ is
\ba\label{} \mc M(\mc
N)&=&\left(
\begin{array}{cc}
\im\mc N+\re\mc N(\im N)^{-1}\re\mc N \ &\  -\re\mc N(\im\mc N)^{-1}
\\&\\ \vspace{-1pt} -(\im\mc N)^{-1}\re\mc N &
(\im\mc N)^{-1}
\end{array}\right)\ .\ea

The indices $\Lambda\,,\Sigma$ of $\mc N_{\Lambda\Sigma}$ are now split as $(0,I)$, according to the decomposition of $4d$ charges with respect to
$5d$ ones, thus $\mc N_{\Lambda\Sigma}$ assumes the block form \ba\label{} {\mc N}_{\Lambda\,\Sigma}&=& \left(
\begin{array}{c|c}
\mc N_{00}\ &\ \mc N_{0\, J}\\
\hline\vspace{3pt}
\mc N_{I\,0}\ & {\mc N}_{I\, J}
\end{array}\right)\ ,
\ea

The kinetic matrix depends on the $70$ scalars of the $\mc N=8$ theory, which are given, in the $5d/4d$ KK reduction, by the $42$ scalars of the
$5d$ theory (encoded in the $5d$ vector kinetic matrix $a_{IJ}=a_{JI}$), by the $27$ axions $a^{I}$ and the dilaton field $e^{\phi}$. In a
normalization that is suitable for comparison to ${\mc N=2}$ , it has the form \ba\label{} \mc N_{\Lambda\Sigma} &=& \left(
\begin{array}{c|c}
&\\
\frac13d -i\left( e^{2\phi}{a}_{IJ}a^{I}a^{J}+e^{6\phi}
\right)\ &\
-\frac12d_{J}+i e^{2\phi}{a}_{KJ}a^{K}
\\&\\
\hline\\ \vspace{-1pt}
-\frac12d_{I}+i e^{2\phi}{a}_{IK}a^{K}\ &
\ d_{IJ}-
ie^{2\phi}{a}_{IJ}\\&\\
\end{array}
\right)\ ,
\ea
where \be\label{short} d\equiv d_{IJK}a^I a^J a^K\quad\ ,\quad
d_{I}\equiv  d_{IJK}a^J a^K\quad\ ,\quad d_{IJ}\equiv  d_{IJK}a^{K}\
. \ee

The black hole potential of \cite{Ceresole:2009jc}, computed from (\ref{Vbh}) using the above formulas, can be rearranged as \ba\label{}
V_{BH}&=&\frac12\left(p^{0}e^{\phi}a^{I} \right)a_{IJ}\left(p^{0}e^{\phi}a^{J} \right)+\frac12\left( p^{0}e^{3\phi}\right)^{2}+\frac12\left(\frac d6 p^{0}e^{-3\phi}\right)^{2}+\nn\\
&&+\frac12\left(\frac12 e^{-\phi}p^{0}d_{I}\right)a^{IJ}\left(\frac12 e^{-\phi}p^{0}d_{J}\right)+\frac12\times 2\left(-p^{0}e^{\phi}a_{I}\right)a_{IJ}\left( p^{J}e^{\phi} \right)+\nn\\
&&+\frac12\times 2\left(\frac d6 p^{0}e^{-3\phi}\right)\left( -\frac12p^{I}d_{I}e^{-3\phi}\right)-\frac12\times 2\left(\frac12\,p^{0}e^{-\phi}d_{I} \right)a^{IJ}\left( p^{K}d_{KJ}e^{-\phi} \right)+\nn\\
&&+\frac12\left( e^{\phi}p^{I}\right)a_{IJ}\left( e^{\phi}p^{J}\right)+\frac12\left( \frac12 e^{-3\phi} p^{K}d_{K}\right)^{2}+\nn\\
&&+\frac12\left(  e^{-\phi}p^{K}d_{KI}\right)a^{IJ}\left( e^{-\phi}p^{L}d_{JL} \right)+
\frac12\times2\left( q_{0}e^{-3\phi} \right)
\left( \frac d6\,p^{0}e^{-3\phi} \right)+ \nn\\
&&+\frac12\times2\left( q_{I}a^{I}e^{-3\phi} \right)\left( \frac d6\,p^{0}e^{-3\phi} \right)+
\frac12\times2\left( q_{I}e^{-\phi} \right)a^{IJ}\left( \frac12p^{0}d_{J}e^{-\phi} \right)+\nn\\
&&-\frac12\times2\left( q_{0}e^{-3\phi} \right)\left( \frac12p^{I}d_{I}e^{-3\phi} \right)-\frac12\times2\left( q_{I}a^{I}e^{-3\phi} \right)\left( \frac12 p^{J}d_{J}e^{-3\phi}\right)+\nn\\
&&-\frac12\times2\left(  q_{I}e^{-\phi}\right)a^{IJ}\left( p^{K}d_{KJ}e^{-\phi} \right)+
\frac12\left( q_{0}e^{-3\phi} \right)^{2}+\frac12\times 2\left( q_{0}e^{-3\phi}\right)\left(q_{I}a^{I}e^{-3\phi} \right)+\nn\\
&&+\frac12\left( q_{I}a^{I}e^{-3\phi} \right)^{2}+\frac12\left( q_{I}e^{-\phi} \right)a^{IJ}\left( q_{J}e^{-\phi} \right)\ ,\nn\\
\ea
with $a^{IJ}=a^{-1}_{IJ}$.
This form shows that it can be written in terms of squares of electric and magnetic components as
\ba\label{} V_{BH}&=&\frac12 (Z^{e}_{0})^{2}+\frac12 \left(Z_{m}^{0}\right)^{2}+\frac12
Z^{e}_{I}a^{IJ}Z^{e}_{J}+\frac12 Z_{m}^{I}a_{IJ}Z_{m}^{J}\ ,
\ea
provided one defines, \ba\label{firstdefCC}
Z^{e}_{0}&=&e^{-3\phi}q_{0}+e^{-3\phi}q_{I}a^{I}+e^{-3\phi}\frac d6\,p^{0}-\frac12e^{-3\phi}p^{I}d_{I}\ ,\nn\\
Z^{0}_{m}&=&e^{3\phi}p^{0}\ ,\nn\\
Z_{I}^{e}&=&\frac12e^{-\phi}p^{0}d_{I}-p^{J}d_{IJ}e^{-\phi}+q_{I}e^{-\phi}\ ,\nn\\
Z^{I}_{m}&=&e^{\phi}p^{I}-e^{\phi}p^{0}a^{I}\ .
\ea
In order to get the symplectic embedding of the four dimensional theory, we still need to complexify the central charges. To this end,
we define the two complex vectors
\ba\label{complexified-charges}
Z_{0}&\equiv&\frac1{\sqrt2}(Z^{e}_{0}+iZ^{0}_{m})\ ,\nn\\
Z_{a}&\equiv&\frac1{\sqrt2}(Z^{e}_{a}+iZ^{a}_{m})\ ,
\ea
where
\begin{eqnarray}\label{}
Z^{e}_{a}&=&Z_{I}^{e}(a^{-1/2})^{I}_{a}\ ,\qquad Z_{m}^{a}=Z_{m}^{I}(a^{1/2})_{I}^{a}
\end{eqnarray}
such that
\ba\label{VBHCC}
V_{BH}&=& | Z_{0}|^{2}+Z_{a}\bar  Z_{a}\ ,
\ea
where now $a=1,...,27$ is a flat index, which can be regarded as a $USp(8)$ antisymmetric traceless matrix.

The potential at the critical point gives the black hole entropy corresponding to the given solution, which in $d=4$ reads
\begin{eqnarray}\label{}
\frac{S_{BH}}\pi&=&\sqrt{|I_{4}|}=V_{BH}^{crit.}\ ,
\end{eqnarray}
while in $d=5$ it is \cite{Andrianopoli:1997hb}
\begin{eqnarray}\label{}
\frac{S_{BH}}\pi&=&3^{3/2}|I_{3}|^{1/2}=\left( 3\,V_{5}^{crit} \right)^{3/4}\ ,
\end{eqnarray}
where $I_{4}$ and $I_{3}$ are the invariants of the $\mc N=8$ theory in $d=4$ and $d=5$ respectively.

\subsection{Symplectic sections}
In virtue of the previous discussion, we can trade the central charge (\ref{ZAB})for the $28$-component vector
\ba\label{}
Z_{A}&=&f^{\Lambda}_{\ph{\Lambda}A}q_{\Lambda}-h_{\Lambda A}p^{\Lambda}\ ,
\ea
where $f$ and $h$ are symplectic sections satisfying the following properties \cite{Gaillard:1981rj,Aschieri:2008ns}
\begin{description}
\item[a)] $\mc N_{\Lambda\Sigma}=h_{\Lambda A}(f^{-1})^{A}_{\ph{A}\Sigma}\ ,$
\item[b)]$i(\mb f^{\dag}\mb h-\mb h^{\dag}\mb f)=\mb{Id}\ ,$
\item[c)]$\mb f^{T}\mb h-\mb h^{T}\mb f=0\ .$
\end{description}
Notice that one still has the freedom of a further transformation
\ba\label{redefM}
h\rar hM\ ,\nn\\
f\rar fM\ ,
\ea
as it leaves invariant the vector kinetic matrix $\mc N$, as well as relations $a) - c)$, when $M$ is a unitary matrix
\ba\label{}
MM^{\dag}=1\ .
\ea
Indeed, when the central charge transforms as
\ba\label{}
Z&\rar& ZM\ ,\nn\\
ZZ^{\dag}&\rar& ZMM^{\dag}Z^{\dag}=ZZ^{\dag}\ ,
\ea
the black hole potential
\ba\label{ZZbar}
V_{BH}\equiv ZZ^{\dag}
\ea
is left invariant.
In our case, we rearrange the 28 indices into a single complex vector index, to be identified, for a suitable choice of $M$, with the two-fold antisymmetric representation of $SU(8)$, according to the decomposition ${\bf 28\rightarrow27+1}$ of $SU(8)\rightarrow USp(8)$; we thus have 
\ba\label{}
Z_{0}&=& f^{\Lambda}_{\ph{\Lambda}0}q_{\Lambda}-h_{\Lambda\,0}p^{\Lambda}=\nn\\
&=& f^{0}_{\ph{0}0}q_{0}+ f^{J}_{\ph{J}0}q_{J}-h_{0\,0}p^{0}-h_{J\,0}p^{J}\ ,\nn\\
Z_{a}&=& f^{\Lambda}_{\ph{\Lambda}a}q_{\Lambda}-h_{\Lambda\,a}p^{\Lambda}=\nn\\
&=& f^{0}_{\ph{0}a}q_{0}+ f^{J}_{\ph{J}a}q_{J}-h_{0\,a}p^{0}-h_{J\,a}p^{J}\ ;\nn\\
\ea
which, from the definition in (\ref{complexified-charges}) yields
\ba\label{tildedZ}
Z_{0}&=&\frac1{\sqrt2}\left[e^{-3\phi}q_{0}+e^{-3\phi}a^{I}q_{I}+\left( e^{-3\phi}\frac d6+ie^{3\phi} \right)p^{0}-\frac12\left(  e^{-3\phi}d_{I} \right)p^{I}\right]\ ,\nn\\
Z_{a}&=&\frac1{\sqrt2}\left[e^{-\phi}q_{I}(a^{-1/2})^{I}_{\ph Ia}+\left(  \frac12e^{-\phi}d_{I}(a^{-1/2})^{I}_{\ph Ia}-ie^{\phi}a^{J}(a^{1/2})^{\ph Ja}_{J}\right)p^{0}+\right.\nn\\
&&\left.\ph{\frac12}-\left( e^{-\phi}d_{IJ}(a^{-1/2})^{I}_{\ph Ia} -ie^{\phi}(a^{1/2})^{\ph Ja}_{J}\right)p^{J}\right]\ .\nn\\
\ea
Thus we consider
\ba\label{}
f^{\Lambda}_{\ph{\Lambda}A}&=&\frac1{\sqrt2}\left(\begin{array}{c|c}
&\\
e^{-3\phi}\ & \ 0 \\&\\
\hline\\ \vspace{-1pt}
e^{-3\phi}a^{I}\ & e^{-\phi}(a^{-1/2})^{I}_{\ph Ia}\\&\\
\end{array}\right)\ ,
\end{eqnarray}
\begin{eqnarray}\label{}
h_{\Lambda\,A}&=&\frac1{\sqrt2}\left(\begin{array}{c|c}
&\\
-e^{-3\phi}\frac d6-ie^{3\phi}\ &\ -\frac12 e^{-\phi}d_{K}(a^{-1/2})^{K}_{\ph Ia}+ie^{\phi}a^{K}(a^{1/2})^{\ph Ka}_{K}\\&\\
\hline\\ \vspace{-1pt}
\frac12 e^{-3\phi}d_{I}\ & e^{-\phi}d_{IJ}(a^{-1/2})^{J}_{\ph Ja}-ie^{\phi}(a^{1/2})^{\ph Ia}_{I}\\&\\
\end{array}\right)\ .
\ea
From ${\mb f}^{-1}$
\ba\label{}
(f^{-1})_{\Lambda}^{\ph{\Lambda}A}&=&\sqrt2\left(\begin{array}{c|c}
&\\
e^{3\phi}\ & \ 0 \\&\\
\hline\\ \vspace{-1pt}
-e^{\phi}a^{I}(a^{1/2})^{\ph Ia}_{I}\ & e^{\phi}(a^{1/2})^{\ph Ia}_{I}\\&\\
\end{array}\right)\ ,
\ea
by matrix multiplication, we find that relations $a)$  $b)$ and $c)$ are fulfilled by $\mb f$ and $\mb h$, that we now recognize to be the symplectic sections.

We finally perform the transformation $f'=fM$ (where $M=f^{-1}f'=h^{-1}h'$), with $M$ unitary matrix, in virtue of identities $a)$, $b)$ and $c)$, valid for both $(f, h)$ and $(f',h')$. A model independent formula for $M$ valid for any $\mc N=2$ d-geometry (in particular, for any truncation of $\mc N=8$ to an $\mc N=2$ geometry, such as the models treated in this paper) is given by the matrix \cite{New}
\begin{eqnarray}\label{}
M=A^{1/2}\hat M G^{-1/2}\ ,
\end{eqnarray}
with 
\begin{eqnarray}\label{}
A=\left(\begin{array}{c|c}
1&0...0\\ 
\hline 
\begin{array}{c}
0\ \\.\ \\.\ \\0\ \end{array}& \ \ \ a_{IJ}\ \ \end{array}
\right)\ ,\qquad  G=\left(\begin{array}{c|c}
1&0...0\\
\hline 
\begin{array}{c}
0\ \\.\ \\.\ \\0\ \end{array}& \ \ \ g_{IJ}\ \ \end{array}
\right)\ ,\qquad g_{IJ}=\frac14e^{-4\phi}a_{IJ}\ ,
\end{eqnarray}
where $\hat M$ is given by
\begin{eqnarray}\label{}
\hat M&=&\frac12
\left(
\begin{array}{cc}
1\ \ &\p_{\bar{J}}K\\
-i\lambda^{I} e^{-2\phi}\ \ & e^{-2\phi}\delta^{I}_{\bar J}+ie^{-2\phi}\lambda^{I}\p_{\bar{J}}K
\end{array}
\right)\ ,
\end{eqnarray}
where ``$-\lambda^{I}$'' are the imaginary parts of the complex moduli $z^{I}=a^{I}-i\lambda^{I}$, and $K$ is the K\"ahler potential $K=-\ln(8\mc V)$, with $\mc V=\frac1{3!}d_{IJK}\lambda^{I}\lambda^{J}\lambda^{K}$; the matrix $\hat M$ satisfies the properties
\begin{eqnarray}\label{}
A\hat MG^{-1}\hat M^{\dag}&=&Id\ ,\nn\\
G^{-1}\hat M^{\dag}A\hat M&=&Id\ .
\end{eqnarray}
For the models considered below, this matrix $M$ does indeed reproduce, for the given special configurations, the formula in eq. (\ref{4dcc}). 

Note that $\hat M$ performs the change of basis between the central charges defined as 
\begin{eqnarray}\label{}
Z_{0}&=&\frac1{\sqrt2}(Z^{e}_{0}+iZ^{0}_{m})\ ,\nn\\
Z_{I}&=&\frac1{\sqrt2}(Z^{e}_{I}+ia_{IJ}Z^{J}_{m})\ ,
\end{eqnarray}
and the special geometry charges $(Z,\,\mc D_{\bar{I}}\overline Z)$, that is the charges in ``curved'' rather than the ``flat'' indices.
%


\section{\label{solutions-5d}Attractors in the 5 dimensional theory}
It was shown in \cite{malda} that the cubic invariant of the five dimensions can be written as
\ba\label{}
I_{3}&=&Z^{\it5}_{1}Z^{\it{5}}_{2}Z^{\it{5}}_{3}\ ,
\ea
where $Z^{\it{5}}_{a}$'s are related to the skew eigenvalues of the $USp(8)$ central charge matrix in the normal frame
\ba\label{Z5}
e_{ab}&=&\left(
\begin{array}{cccc}
Z^{\it{5}}_{1}+Z^{\it{5}}_{2}-Z^{\it{5}}_{3}&0&0&0\\
0&Z^{\it{5}}_{1}+Z^{\it{5}}_{3}-Z^{\it{5}}_{2}&0&0\\
0&0&Z^{\it{5}}_{2}+Z^{\it{5}}_{3}-Z^{\it{5}}_{1}&0\\
0&0&0&-(Z^{\it{5}}_{1}+Z^{\it{5}}_{2}+Z^{\it{5}}_{3})
\end{array}
\right)\otimes\left(\begin{array}{cc}0&\ 1\\-1&\ 0\end{array}\right)\ .\nn\\
\ea
We consider a configuration of only three non-vanishing electric charges $(q_{1},q_{2},q_{3})$, that we can take all non-negative. We further
confine to two moduli $\lambda_{1}, \lambda_{2}$, describing a geodesic submanifold $SO(1,1)^{2}\in E_{6(6)}/USp(8)$ whose special geometry is determined by the constraint\ba\label{}
\frac1{3!}d_{IJK}\hat\lambda^{I}\hat\lambda^{J}\hat\lambda^{K}&=&\hat\lambda^{1}\hat\lambda^{2}\hat\lambda^{3}=1\ ,
\ea
where $\hat\lambda^{I}=\mc V^{-1/3}\lambda^{I}$,  defining the $stu-$model \cite{Ceresole:2007rq}. 

The metric $a_{IJ}$, restricted to this surface, takes the diagonal form
\be\label{metrica5d}
a_{IJ}=-\frac{\p^{2}}{\p\hat\lambda^{I}\p\hat\lambda^{J}}\log \mc V\big|_{\mc V=1}
=\left(
\begin{array}{ccc}
\frac1{\hat\lambda_{1}^{2}}&0&0\\
0&\frac1{\hat\lambda_{2}^{2}}&0\\
0&0&\frac1{\hat\lambda_{3}^{2}}=\hat\lambda_{1}^{2}\hat\lambda_{2}^{2}\\
\end{array}
\right)\ ,
\ee
and the five dimensional black hole potential for electric charges is\footnote{In an analogous way, the black hole potential for magnetic charges, $V_{5}^{m}=\sum_{a=1}^{3}Z_{a}^{\it5}(p)Z_{a}^{\it5}(p)$, is obtained by replacing $q_{I}\rar p^{I}$ and $a^{IJ}\rar a_{IJ}$ \cite{Ceresole:2007rq,Andrianopoli:1997hb}, with $Z_{a}^{\it5}(p)=p^{I}(a^{1/2})^{\ph Ia}_{I}$ .}
\ba\label{V5}
V^{e}_{5}&=&q_{I}a^{IJ}q_{J}=\sum_{a=1}^{3}Z^{\it5}_{a}(q)Z^{\it5}_{a}(q)\ ,
\ea
with $Z^{\it5}_{a}(q)=(a^{-1/2})^{I}_{\ph Ia}\,q_{I}$; the moduli
at the attractor point of the 5-dimensional solution are (see eq. 4.4 and 4.7 of \cite{Ceresole:2007rq})
\ba\label{moduliAttr}
\hat\lambda^{I}_{\,crit}&=&\frac{I_{3}^{1/3}}{q^{I}}\ ,
\ea
and
\ba\label{}
V_{5}^{crit}&=&3|q_{1}q_{2}q_{3}|^{2/3}=3I_{3}^{2/3}\ ,\nn\\
a^{IJ}_{crit}&=&\frac{I_{3}^{2/3}}{q_{I}^{2}}\delta^{IJ}
\ea
with no sum over repeated indices. We find
\ba\label{}
Z_{a}^{{\it5},\,crit}&=I_{3}^{1/3}\ ,\qquad\
I_{3}=Z^{\it5}_{1}Z^{\it5}_{2}Z^{\it5}_{3}\ .
\ea
These relations also allow to connect the potential in (\ref{V5})
\ba\label{} 
V_{5}&=&(Z^{\it5}_{1})^{2}+(Z_{2}^{\it5})^{2}+(Z_{3}^{\it5})^{2}\ ,
\ea
with the form given in terms of the central charges  \cite{Andrianopoli:1997hb}, where it is the trace of the square matrix
\ba\label{}
V_{5}&=&\frac12Z^{\it5}_{ab}Z^{{\it5}\,ab}\ .
\ea
The eigenvalues of $Z^{\it5}_{ab}$ are written in (\ref{Z5}) in terms of $Z^{\it5}_{1},Z^{\it5}_{2},Z^{\it5}_{3}$.
The 5d central charge matrix in the normal frame at the attractor point thus becomes
\ba\label{}
e_{ab}
&=&
\left(\begin{array}{cccc}
I_{3}^{1/3}\epsilon&0&0&0\\
0&I_{3}^{1/3}\epsilon&0&0\\
0&0&I_{3}^{1/3}\epsilon&0\\
0&0&0&-3I_{3}^{1/3}\epsilon
\end{array}\right)\ ,
\ea
which shows the breaking  $USp(8)\rar USp(6)\times USp(2)$.
\section{\label{solutions-4d}Attractors in the 4 dimensional theory}
In this section we reconsider the attractor solutions  found in \cite{Ceresole:2009jc,Ceresole:2007rq}and we
reformulate them in terms of the present formalism based on central charges. We separately examine the three ``axion free" configurations.

\subsection{\label{electr. sol.}Electric solution $Q=(p^{0}\,,q_{i})$}

Let us first compute the 4dim central charge for the electric charge configuration with vanishing axions; using (\ref{tildedZ}) we find
\begin{eqnarray}\label{}
Z_{0}=\frac i{\sqrt2}e^{3\phi}p^{0}\ ,\qquad
Z_{a}=\frac1{\sqrt2}e^{-\phi}q_{I}(a^{-1/2})^{I}_{\ph Ia}\ .
\end{eqnarray}
The 4-dim potential is
\ba\label{}
V_{BH}&=&\frac12 e^{-2\phi}V^{e}_{5}+\frac12e^{6\phi}(p^{0})^{2}\ ,
\ea
(where $\phi$ is connected to the volume used in ref.\cite{Ceresole:2007rq} by the formula $\mc V=e^{6\phi}$)
and has the same critical points of the 5 dimensional potential, since
\ba\label{}
\frac{\p V_{BH}}{\p\lambda^{I}}=0\iff\frac{\p V^{e}_{5}}{\p\hat\lambda^{I}}=0\ ,\qquad \forall\  I=1,2\, .
\ea
The attractor values of $\hat\lambda^{I}$ are still given by (\ref{moduliAttr}),
while the $\phi$ field at the critical point is \cite{Ceresole:2007rq}
\ba\label{}
e^{8\phi}|_{crit.}&=&I_{3}^{2/3}(p^{0})^{-2}\ .
\ea
This fixes the central charges at the attractor point to be
\ba\label{eCC}
Z_{0}^{\ph 0attr}&=&\frac i{\sqrt2}|p^{0}q_{1}q_{2}q_{3}|^{1/4}sign(p^{0})=\frac i2|I_{4}|^{1/4}sign(p^{0})\ ,\nn\\
Z_{a}^{\ph{I}attr}&=&\frac1{\sqrt2}I_{3}^{-1/12}(p^{0})^{1/4}q_{I}\frac{I_{3}^{1/3}}{q_{I}}=\frac12|I_{4}|^{1/4}\ ,
\ea
where the quartic invariant is $I_{4}=-4\,p^{0}q_{1}q_{2}q_{3}$.  So we find
\ba\label{CC}
Z^{crit}_{1}= Z^{crit}_{2}= Z^{crit}_{3}=\frac12|I_{4}|^{1/4}\equiv Z\ ,\qquad Z^{crit}_{0}=\frac i2|I_{4}|^{1/4}sign(p^{0})\equiv iZ_{0}\ .
\ea
Let us define the 4d central charge matrix as \ba\label{4dcc}
2Z_{AB}=e_{AB}-i Z^{0}\Omega\ ,
\ea
where $e_{AB}$ is the matrix in (\ref{Z5}) in which, instead of $Z^{\it5}_{1},Z^{\it5}_{2},Z^{\it5}_{3}$ of the 5d theory,  we now write the 4d $Z_{a}$'s defined in (\ref{tildedZ}).
it can be readily seen that for axion free solutions eq. (\ref{4dcc}) correctly gives 
\begin{eqnarray}\label{}
V_{BH}=\sum_{i}|z_{i}|^{2}=|Z_{0}|^{2}+\sum_{a}|Z_{a}|^{2}
\end{eqnarray} where $z_{i}$'s, for $i=1,..,4$, are the (complex skew-diagonal) elements of $Z_{AB}$.
We then have
\ba\label{}
2Z_{AB}&=&\left(\begin{array}{cccc}
Z\epsilon&0&0&0\\
0&Z\epsilon&0&0\\
0&0&Z\epsilon&0\\
0&0&0&-3Z\epsilon
\end{array}\right)+\left(\begin{array}{cccc}
Z_{0}\epsilon&0&0&0\\
0&Z_{0}\epsilon&0&0\\
0&0&Z_{0}\epsilon&0\\
0&0&0&Z_{0}\epsilon
\end{array}\right)= \nn\\ \nn\\ \nn\\
&=&\left(\begin{array}{cccc}
(Z+Z_{0})\epsilon&0&0&0\\
0&(Z+Z_{0})\epsilon&0&0\\
0&0&(Z+Z_{0})\epsilon&0\\
0&0&0&(-3Z+Z_{0})\epsilon
\end{array}\right)\ .\nn\\
\ea
Since (\ref{eCC}) and (\ref{CC}) yield that $Z=|Z_{0}|$, depending on the choice $p^{0}>0$ or $p^{0}<0$, two different solutions arise. In fact,
\ba\label{electrBPS}
Z+Z_{0}=0\quad \rar\quad Z_{AB}=\left(\begin{array}{cccc}
0&0&0&0\\
0&0&0&0\\
0&0&0&0\\
0&0&0&2Z_{0}
\end{array}\right)\otimes\epsilon\ ,
\ea
gives the $\frac18$-BPS solution when $p^{0}<0$ and shows $SU(6)\times SU(2)$ symmetry.  Conversely,
\ba\label{electrnonBPS}
Z=Z_{0}\quad \rar\quad Z_{AB}=\left(\begin{array}{cccc}
Z_{0}&0&0&0\\
0&Z_{0}&0&0\\
0&0&Z_{0}&0\\
0&0&0&-Z_{0}
\end{array}\right)\otimes\epsilon\ ,
\ea
is the non-BPS solution that corresponds to the choice $p^{0}>0$, with residual $USp(8)$ symmetry.
\subsection{Magnetic solution $Q=(p_{i}\,,q^{0})$}

This case is symmetric to the electric solution of Section \ref{electr. sol.}. If we take all positive magnetic charges, then the cubic invariant is
$
I_{3}=p^{1}p^{2}p^{3}\ ,
$
the quartic invariant is $I_{4}=4\,q_{0}\,p^{1}p^{2}p^{3}$ and the values of the critical 5d moduli are now (see eq. (5.3) of \cite{Ceresole:2007rq})
\ba\label{}
\hat\lambda^I=\frac{p^{I}}{I_{3}^{1/3}}\ .
\ea
The central charges for this configuration are, from (\ref{tildedZ}),
\begin{eqnarray}\label{}
Z_{0}=\frac1{\sqrt2}e^{-3\phi}q_{0}\ ,\qquad \quad Z_{a}=\frac i{\sqrt2}e^{\phi}p^{I}(a^{1/2})_{I}^{\ph Ia}\ ,
\end{eqnarray}
and the black hole potential is
\ba\label{}
V_{BH}&=&\frac12e^{2\phi}V^{m}_{5}+\frac12e^{-6\phi}(q_{0})^{2}\ .
\ea
This gives the attractor value of the $\phi$ field as
\ba\label{}
e^{8\phi}|_{crit.}&=&I_{3}^{-2/3}(q_{0})^{2}\ .
\ea
At the attractor point $(a_{crit.}^{1/2})_{IJ}=(\hat\lambda^{I})^{-1}{\delta_{IJ}}$, and the magnetic central charges are
\ba\label{}
Z^{crit}_{a}&=&\frac i{\sqrt2}(I_{3})^{1/4}|q_{0}|^{1/4}=\frac i{2}|I_{4}|^{1/4}\equiv iZ\ ,\qquad\ a=1,2,3\ .
\ea
We can then write the central charge matrix corresponding to the $\mb 27$ representation in the normal frame as
\ba\label{}
e_{AB}=\left(\begin{array}{cccc}
Z\epsilon&0&0&0\\
0&Z\epsilon&0&0\\
0&0&Z\epsilon&0\\
0&0&0&-3Z\epsilon
\end{array}\right)\ .
\ea
To describe the four dimensional solution we need the electric central charge, that at the attractor point is
\ba\label{}
Z_{0}^{crit}&=&\frac1{\sqrt2}(I_{3})^{1/4}|q_{0}|^{1/4}\,sign(q_{0})=\frac1{2}|I_{4}|^{1/4}\,sign(q_{0})\equiv Z_{0}\ .\nn
\ea
Then, using the definition(\ref{4dcc}) the complete 4d central charge matrix is
\ba\label{}
2Z_{AB}&=&i\left(\begin{array}{cccc}
Z\epsilon&0&0&0\\
0&Z\epsilon&0&0\\
0&0&Z\epsilon&0\\
0&0&0&-3Z\epsilon
\end{array}\right)-i\left(\begin{array}{cccc}
Z_{0}\epsilon&0&0&0\\
0&Z_{0}\epsilon&0&0\\
0&0&Z_{0}\epsilon&0\\
0&0&0&Z_{0}\epsilon
\end{array}\right)= \nn\\ \nn\\ \nn\\
&=&e^{i\pi/2}\left(\begin{array}{cccc}
(Z-Z_{0})\epsilon&0&0&0\\
0&(Z-Z_{0})\epsilon&0&0\\
0&0&(Z-Z_{0})\epsilon&0\\
0&0&0&(-3Z-Z_{0})\epsilon
\end{array}\right)\ .\nn\\
\ea
The $sign(q_{0})$ determines whether the solution is supersymmetric or not. We may have
\ba\label{}
q_{0}>0\quad\rar\quad Z=Z_{0}\ ,\nn\\ \nn\\
Z_{AB}=e^{i\pi/2}\left(\begin{array}{cccc}
0&0&0&0\\
0&0&0&0\\
0&0&0&0\\
0&0&0&-2Z_{0}
\end{array}\right)\otimes \epsilon
\ea
which is a magnetic $\frac1{8}$-BPS solutions with $SU(6)\times SU(2)$ symmetry, or
\ba\label{magnnonBPS}
q_{0}<0\quad\rar\quad Z=-Z_{0}\ ,\ \ \ \ \ \ \ \ \nn\\ \nn\\
Z_{AB}=e^{i\pi/2}\left(\begin{array}{cccc}
-Z_{0}&0&0&0\\
0&-Z_{0}&0&0\\
0&0&-Z_{0}&0\\
0&0&0&Z_{0}
\end{array}\right)\otimes \epsilon
\ea
which is the non-BPS solution with  $USp(8)$ symmetry. These solutions have the same $Z_{0}$ as the electric ones, but now the choice of positive $q_{0}$ charge leads to the supersymmetric solution while the negative $q_{0}$ charge gives the non-supersymmetric one, in contrast with what happened for the choice of $p^{0}$ in the electric case in eq. (\ref{electrBPS}) and (\ref{electrnonBPS}).
\subsection{KK dyonic solution $Q=(p^{0}\,,q_{0})$}
This charge configuration also has vanishing axions, and the only non-zero charges give
\ba\label{}\begin{array}{c}
Z^{e}_{0}=e^{-3\phi}q_{0}\ ,\quad
Z^{0}_{m}=e^{3\phi}p^{0}\ ,\\
\dar\\
Z_{0}=\frac1{\sqrt2}(e^{-3\phi}q_{0}+ie^{3\phi}p^{0})\ .
\end{array}
\ea
Since none of the 5 dimensional charges are turned on, the four dimensional black hole potential is
\ba\label{}
V_{BH}&=&\frac12\left[
e^{-6\phi}q_{0}^{2}+e^{6\phi}(p^{0})^{2}
\right]\ ,
\ea
which is extremized at the horizon by the value of the $\phi$ field
\ba\label{}
e^{6\phi}|_{crit.}&=&\left|\frac{q_{0}}{p^{0}}\right|\ .
\ea
We only focus on the case $p^{0}>0$ and $q_{0}>0$, since all the other choices are related to this by a duality rotation.
Evaluating the central charge at the attractor point we find
\ba\label{}
Z_{0}^{crit}&=&\sqrt{|p^{0}q_{0}|}\frac{1+i}{\sqrt2}=\sqrt{|p^{0}q_{0}|}e^{i\pi/4}\ .
\ea
Following the prescription in (\ref{4dcc}) we find that at the attractor point
\ba\label{KKnonBPS}
2Z_{AB}&=&-i Z_{0}\Omega=\nn\\
&=&-ie^{i\pi/4}\left(\begin{array}{cccc}
\sqrt{|p^{0}q_{0}|}\epsilon&0&0&0\\
0&\sqrt{|p^{0}q_{0}|}\epsilon&0&0\\
0&0&\sqrt{|p^{0}q_{0}|}\epsilon&0\\
0&0&0&\sqrt{|p^{0}q_{0}|}\epsilon
\end{array}\right)
\ea
that gives a non-BPS 4 dimensional black hole with $I_{4}=-(p^{0}q_{0})^{2}$.

Note that eqs. (\ref{electrnonBPS}), (\ref{magnnonBPS}) and (\ref{KKnonBPS}) imply that the sum of the phases of the four complex skew entries is $\pi$, as appropriate to a non-BPS $\mc N=8$ solution \cite{Ferrara:2006em}. Also, in all cases, $V_{BH}|_{crit.}=\sqrt{|I_{4}|}$.
\subsection{\label{N8N2}$\mc N=8$ and $\mc N=2$ attractive orbits at $d=5$ and $d=4$}
We now compare the different interpretations in the $\mc N=8$ and $\mc N=2$ theories of the critical points of the very same black hole $4d$ potential, in terms of the axion-free electric solution (sec. \ref{electr. sol.}) as discussed in this paper and in ref. \cite{Ceresole:2007rq}.

Since the ``normal frame'' solution is common to all symmetric spaces (with rank three), it can be regarded as the generating solution of any model. So we confine our attention to the exceptional $\mc N=2$ (octonionic) $E_{7(-25)}$ model  \cite{Gunaydin:1983bi}  which has a charge vector in $5d$ and $4d$ of the same dimension as in $\mc N=8$ supergravity. At $d=5$ the duality group is $E_{6(-26)}$, with moduli space of vector multiplets $E_{6(-26)}/F_{4}$.

It is known \cite{gunaydin1,Ferrara:2006xx} that in $d=5$ there are two different charge orbits,
\begin{eqnarray}\label{}
\mc O_{d=5,\,BPS}^{\mc N=2}&=&\frac{E_{6(-26)}}{F_{4}}\ ,
\end{eqnarray}
the BPS one, and the non BPS one
\begin{eqnarray}\label{}
\mc O_{d=5,\,non-BPS}^{\mc N=2}&=&\frac{E_{6(-26)}}{F_{4(-20)}}\ ,
\end{eqnarray}
The latter one precisely corresponds to the non supersymmetric solution and to $(++~-)$, $(--~+)$ signs of the $q_{1},q_{2},q_{3},$ charges (implying $\p{Z}\neq0$). For charges of the same sign $(+++)$, $(---)$ one has the $\frac18$BPS solution ($\p Z=0$), as discussed in \cite{Ceresole:2007rq}.

It is easy to see that in the $\mc N=8$ theory all these solutions just interchange $Z_{1},Z_{2},Z_{3}$ and $Z_{4}=-3Z_{3}$ but always give a normal frame matrix of the form
\begin{eqnarray}\label{}
Z_{ab}&=&\left(\begin{array}{cccc}
Z\epsilon&0&0&0\\
0&Z\epsilon&0&0\\
0&0&Z\epsilon&0\\
0&0&0&-3Z\epsilon
\end{array}\right)\ ,
\end{eqnarray}
which has $USp(6)\times USp(2)\in F_{4(4)}$ as maximal symmetry. Another related observation is that while $E_{6(-26)}$ contains both $F_{4}$ and $F_{4(-20)}$, so that one expects two orbits and two classes of solution, in the $\mc N=8$ case $E_{6(6)}$ contains only the non compact $F_{4(4)}$, thus only one class of solutions is possible.

These orbits and critical points at $d=5$ have a further story when used to study the $d=4$ critical points with axion free solutions as it is the case for the electric $(p^{0},q_{1},q_{2},q_{3})$ configuration. Since in this case $I_{4}=-4p^{0}q_{1}q_{2}q_{3}$, in the $\mc N=8$ case, once one choose $q_{1},\,q_{2},\,q_{3}>0$, the $I_{4}>0$, $p^{0}<0$ solution is BPS, while the $I_{4}<0$, $p^{0}>0$ is non BPS.

Things again change in $\mc N=2$ \cite{Bellucci:2006xz}, when now we consider the solution embedded in the Octonionic model with $4d$ moduli space $E_{7(-25)}/E_{6}\times U(1)$. A new non BPS orbit in $d=4$ is generated, corresponding to $Z=0$ ($\p Z\neq0$) solution, so three $4d$ orbits exist in this case depending whether the $(+++)$ and $(++-)$ solutions are combined with $-p^{0}\lessgtr0$.
So
\begin{eqnarray}\label{}
&(+,+++)&\quad \textrm{is BPS with }I_{4}>0\ ,\ \ \ \mc O=\frac{E_{7(-25)}}{E_{6}}\ ,\\
&(-,-++)&\quad \textrm{is non BPS with }I_{4}>0\ ,\ \ \ \mc O=\frac{E_{7(-25)}}{E_{6(-14)}}\ ,\\
&(+,-++)&\ \textrm{or}\ (-,+++)\quad \textrm{is non BPS with }I_{4}<0\ ,\ \ \ \mc O=\frac{E_{7(-25)}}{E_{6(-26)}}\ .\quad
\end{eqnarray}

\section{Maurer-Cartan equations of the four dimensional theory}
Let us call Maurer-Cartan equations\cite{revisited} those which give the derivative of the central charges
(coset representatives) with respect to the moduli $\phi,\ a^{I},\ \lambda^{i}$. Using (\ref{firstdefCC}) we have
\ba\label{}
\p_{\phi}Z^{e}_{0}=-3Z^{e}_{0}\ ,\qquad \p_{\phi}Z_{m}^{0}=3Z^{0}_{m}\ ,\nn\\
\p_{\phi}Z_{I}^{e}=-Z^{e}_{I}\ ,\qquad\p_{\phi}Z^{I}_{m}=Z^{I}_{m}\ ,
\ea
and
\ba\label{}
\frac{\p Z_{0}^{e}}{\p a^{I}}=e^{-2\phi}Z_{I}^{e}\ ,&\qquad& \frac{\p Z^{0}_{m}}{\p a^{I}}=0\ ,\nn\\
\frac{\p Z_{m}^{I}}{\p a^{J}}=-\de_{J}^{I}e^{-2\phi}Z^{0}_{e}\ ,&\qquad&\frac{\p Z_{I}^{e}}{\p a^{J}}=-e^{-2\phi}d_{IJK}Z_{m}^{K}\ .
\ea
In our notation the 5d metric  $a_{IJ}$, ($I,J=1,..,27$) can also be rewritten with a pair of antisymmetric (traceless) indices
\ba\label{}
a_{\Lambda\Sigma\,,\Delta\Gamma}=L^{ab}_{\ph{ab}\Lambda\Sigma}L_{\Delta\Gamma\,ab}\ ,
\ea
where $L^{ab}_{\ph{ab}\Lambda\Sigma}$ is the coset representative; in a fixed gauge (where $a,b$ and $\Lambda,\Sigma$ indices are identified)
\ba\label{}
L_{I}^{\ph Ia}&=&(a^{1/2})_{I}^{\ph Ia}\ ,\qquad (\bar L_{Ia}=L^{T}_{Ia})
\ea
The object $\bb P_{i}\equiv a^{1/2}\p_{i}a^{-1/2}$ can be regarded as the Maurer-Cartan connection (see reference \cite{Sezgin:1981ac}).
In fact, by reminding that $Z^{e}_{a}=Z_{I}^{e}(a^{-1/2})^{I}_{\ph Ia}$, we have $\p_{i}Z^{e}_{a}=(\p_{i}a^{-1/2})^{I}_{\ph Ia}Z_{I}^{e}$ (
since $\p_{i}Z^{e}_{I}=0$).  Since we can also write
\ba\label{}
\p_{i}Z^{e}_{a}=(\p_{i}a^{-1/2})^{I}_{\ph Ia}(a^{1/2})^{\ph Ib}_{I}Z_{b}^{e}
\ea
we find that $\bb P_{i,a}^{\ph{i.a}b}$ is such that
\ba\label{}
\p_{i}Z^{e}_{a}&=&\bb P_{i,a}^{\ph{i.a}b}Z^{e}_{b}\ .
\ea
Notice that using
$
\bb P_{i,a}^{\ph{i,a}b}=Q_{i,a}^{\ph{i,a}b}+V_{i,a}^{\ph{i,a}b}\ ,
$
we identify a connection which satisfies
\ba\label{}
\nabla_{i}Z_{a}^{e}&=&V_{a}^{\ph ab}Z_{b}^{e}\ ,
\ea
with $\nabla_{i}=\p_{i}-Q_{i}$.

\subsection{Attractor equations from Maurer-Cartan equations}
We can now use this formalism to write the attractor equations for
the potential
\ba\label{}
V_{BH}&=&\frac12 (Z_{0}^{e})^{2}+\frac12 (Z^{0}_{m})^{2}+\frac12Z_{I}^{e}a^{IJ}Z_{J}^{e}+\frac12Z_{m}^{I}a_{IJ}Z^{J}_{m}\ .
\ea
By differentiating with respect to $\phi,\ a^{I},\ \lambda^{i}$, we get
\ba\label{critphi}
\p_{\phi}V_{BH}&=&-3(Z_{0}^{e})^{2}+3(Z^{0}_{m})^{2}-Z_{I}^{e}a^{IJ}Z_{J}^{e}+Z_{m}^{I}a_{IJ}Z^{J}_{m}=0\ ,\\
\label{critaxions}
\p_{a^{I}}V_{BH}&=&e^{-2\phi}\left[  Z_{0}^{e}Z_{I}^{e}-Z_{J}^{e}a^{JK}d_{IKL}Z^{L}_{m}-Z^{0}_{m}a_{IJ}Z^{J}_{m} \right]=0\ ,\\
\p_{\lambda^{i}}V_{BH}&\equiv&\p_{i}V_{BH}=\frac12Z_{I}^{e}\,\p_{i}a^{IJ}\,Z^{e}_{J}+\frac12Z_{m}^{I}\,\p_{i}a_{IJ}\,Z^{J}_{m}=0\ .
\ea
From (\ref{critaxions}) we see that a solution with $a^{I}=0$ implies
\ba\label{}
\p_{a^{I}}V_{BH}\big|_{a^{I}=0}=0=e^{-2\phi}\left[e^{-4\phi}q_{0}q_{I}-q_{J}a^{JK}d_{IKL}p^{L}-e^{4\phi}p^{0}a_{IJ}p^{J}\right]=0\ ,
\ea
which is trivially satisfied if we set $\neq 0$ $(q_{0},p^{0})$ or $(q_{0}, p^{I})$ or $(p^{0},q_{I})$.

From (\ref{critphi}) we see that for an axion-free solution, if $Z^{e}_{0},Z^{I}_{m}=0$, we get
\ba\label{}
3(Z^{0}_{m})^{2}=Z_{I}^{e}a^{IJ}Z^{e}_{J}\ ,
\ea
and if $a_{IJ}$ is diagonal, $I=J=1,2,3$, we obtain
\ba\label{solFromAlgebrEq}
3(Z^{0}_{m})^{2}&=&(Z^{e}_{1})^{2}a^{11}+(Z^{e}_{2})^{2}a^{22}+(Z^{e}_{3})^{2}a^{33}\ ,
\ea
which is compatible with $Z^{e}_{1}=Z^{e}_{2}=Z^{e}_{3}=\pm Z^{0}_{m}$\ .

The derivative with respect to the 5d moduli $\lambda^{i}$, $i=1,..,42$ for $\mc N=8$ theory, only receives contributions from
the matrix $a_{IJ}$. Indeed since $Z_{I}^{e}$, $Z_{m}^{I}$ do not depend on the $\lambda^{i}$(see eq.\ref{firstdefCC}), one finds
\ba\label{critmoduli2}
\p_{i}V_{4}&=&0=Z_{I}^{e}\,\p_{i}a^{IJ}\,Z^{e}_{J}+Z_{m}^{I}\,\p_{i}a_{IJ}\,Z^{J}_{m}\ .
\ea
By rewriting the charges multiplied by $(a^{-1/2})^{I}_{\ph Ia}$ and $(a^{1/2})^{\ph Ia}_{I}$ so that
\ba\label{}
Z_{a}^{e}&\equiv&Z_{I}^{e}\,(a^{-1/2})^{I}_{\ph{I}a}\ ,\qquad Z^{a}_{m}=Z_{m}^{I}(a^{1/2})^{a}_{\ph{a}I}\ ,
\ea
we have
\ba\label{}
\p_{i}Z_{a}^{e}&=&\bb P_{i,a}^{\ph{i.a}b}Z_{b}^{e}\ ,\qquad
\bb P_{i,a}^{\ph{i.a}b}=\p_{i}(a^{-1/2})^{I}_{\ph{I}a}(a^{1/2})^{\ph Ib}_{I}\ ,\nn\\
\p_{i}Z^{a}_{m}&=&\bb P_{i\ph{a}b}^{\ph{i}a}Z_{m}^{b}\ ,\qquad
\bb P_{i\ph ab}^{\ph ia}=\p_{i}(a^{1/2})^{\ph Ia}_{I}(a^{-1/2})^{I}_{\ph Ib}\ ,
\ea
where $\bb P_{i\ph{a}b}^{a}=-\bb P_{i\,b}^{\ph{i\,b}a}$ since
$
\p_{i}(Z_{a}^{e}Z^{a}_{m})=0\ .
$
Then we also have
\ba\label{}
\p_{i}(Z^{e}_{a}Z^{e}_{a})&=&Z^{e}_{a}(\bb P_{ia}^{\ph{ia}b})Z^{e}_{b}=\nn\\
&=&Z^{e}_{a}\bb P_{i,ab}Z^{e}_{b}=\nn\\
&=&Z^{e}_{a}\bb P_{i\,(ab)}Z^{e}_{b}=0\ ,
\ea
and if we split $\bb P_{i,ab}=Q_{i\,[ab]}+V_{i\,(ab)}$, with
\begin{eqnarray}\label{}
\bb P_{i\ph ab}^{\ph ia}&=&Q_{i\ph ab}^{\ph ia}+V_{i\ph ab}^{\ph ia}\ ,\nn\\
\bb P_{i,a}^{\ph{i.a}b}&=&Q_{i,a}^{\ph{i.a}b}-V_{i,a}^{\ph{i.a}b}\ ,
\end{eqnarray}
the critical condition implies
\ba\label{}
\p_{i}(Z^{e}Z^{e})=Z_{a}^{e}V_{i\,(ab)}Z^{e}_{b}=0\ ,
\ea
and the analogue equation for magnetic charges
\ba\label{}
\p_{i}(Z^{m}Z^{m})&=&Z_{m}^{a}V_{i\,(ab)}Z_{m}^{b}=0\ ,
\ea
so that only the vielbein $V_{i\,,ab}$ enters in the equations of motion.

The criticality condition on the potential of eq.\,(\ref{critmoduli2}) now gives
\ba\label{}
\p_{i}V_{BH}=0\quad\rar\quad Z_{a}^{e}V_{i}^{\ph{i}ab}Z^{e}_{b}+Z_{m}^{a}V_{i,\,ab}Z_{m}^{b}=0\ ,
\ea
thus, for electric configurations ($Z_{m}^{b}=0$) with $a^{I}=0$,
\ba\label{}
Z^{e}_{a}V_{i}^{\ph iab}Z_{b}^{e}=0\ .
\ea
Comparing results of \cite{Andrianopoli:1997hb} with our formul\ae\  we see that
$V_{1},\ V_{2},\ V_{3},$ with $V_{1}+V_{2}+V_{3}=0$, in the case where the metric $a_{IJ}$ is diagonal, correspond to
\ba\label{}
(a^{-1/2})^{I}_{\ph Ia}\p_{i}(a^{1/2})_{J}^{\ph Ja}=(a^{-1/2})^{I}\p_{i}(a^{1/2})_{I}=\bb P_{i\ph II}^{\ph iI}=V_{i\ph II}^{\ph iI}\equiv V_{i}^{\ph iI}\ ,
\ea
where $(a^{-1/2})^{I}_{\ph II}\equiv (a^{-1/2})^{I}$, 
$(a^{1/2})^{\ph II}_{I}\equiv (a^{1/2})_{I}$,
$I=1,2,3$, and using (\ref{metrica5d}) we find
\ba\label{}
V_{1}^{I}&=&\left(\frac1{\hat\lambda_{1}}\,,0\,,-\frac1{\hat\lambda_{1}}\right)\ ,\nn\\
V_{2}^{I}&=&\left( 0\,,\frac1{\hat\lambda_{2}}\,,-\frac1{\hat\lambda_{2}} \right)\ .
\ea Indeed,
\ba\label{}
\sum_{i=1,2,3} V_{i}^{I}=0\ ,
\ea
so, by using eq. (2.31)-(2.33) of ref. \cite{Andrianopoli:1997hb} one gets the desired result.
In fact, using the definitions of $\bb P_{1}^{I}$ and $\bb P_{2}^{I}$ we get from the $\hat\lambda^{i}$ equations of motion
\ba\label{}
\sum_{I}Z^{e}_{I}V^{I}_{i}Z^{e}_{I}&=&0\ ,
\ea
which explicitly gives
\ba\label{}
Z_{1}^{e}Z_{1}^{e}-Z_{3}^{e}Z_{3}^{e}&=&0\ ,\nn\\
Z_{2}^{e}Z_{2}^{e}-Z_{3}^{e}Z_{3}^{e}&=&0\ ,
\ea
whose solution, combined with eq. (\ref{solFromAlgebrEq}), gives
\ba\label{}
&(Z_{1}^{e})^{2}=(Z_{2}^{e})^{2}=(Z_{3}^{e})^{2}&=(Z^{0}_{m})^{2}\ ,\nn\\
&\dar&\nn\\
&Z_{1}^{e}=Z_{2}^{e}=Z_{3}^{e}=\pm Z^{0}_{m}\ ,&
\ea
all the other sign choices being equivalent in the 5d theory.

\section*{Acknowledgments}

\noindent A.C. and A.G. are glad to thank the CERN Theory Group for the kind hospitality during the final stages of this investigation.
This work is supported in part by the ERC Advanced Grant no. 226455, \textit{``Supersymmetry, Quantum Gravity and Gauge Fields''}
(\textit{SUPERFIELDS}). The work of A.~C.~is partially supported by MIUR-PRIN contract 20075ATT78, while
the work of S.~F.~is partially supported in part by D.O.E.~grant DE-FG03-91ER40662, Task C.


\end{document}